# Orbital-angular-momentum-resolved electron magnetic chiral dichroism

*Enzo Rotunno[1], Matteo Zanfrognini[1], Stefano Frabboni[1,2], Jan Rusz[3], Rafal E. Dunin Borkowski[4], Ebrahim Karimi[5] and Vincenzo Grillo[1]*

1. CNR-NANO via G Campi 213/a, 41125 Modena, Italy
2. Dipartimento FIM, Università di Modena e Reggio Emilia, via G. Campi 213/a, 41125, Modena, Italy
3. Department of Physics and Astronomy, Uppsala University, P.O. Box 516, 75120 Uppsala, Sweden
4. Ernst Ruska-Centre for Microscopy and Spectroscopy with Electrons and Peter Grünberg Institute, Forschungszentrum Jülich, 52425 Jülich, Germany
5. Department of Physics, University of Ottawa, 150 Louis Pasteur, Ottawa, Ontario K1N 6N5, Canada

**Abstract**

We propose to use the recently introduced orbital angular momentum spectrometer in a transmission electron microscope to perform electron magnetic chiral dichroism experiments, dispersing the inelastically scattered electrons from a magnetic material in both energy and angular momentum. The technique offers several advantages over previous formulations of electron magnetic chiral dichroism as it requires much simpler experimental conditions in terms of specimen orientation and thickness.
A novel simulation algorithm, based on the multislice description of the beam propagation, is used to anticipate the advantages of the new approach over current electron magnetic chiral dichroism implementations. Numerical calculations confirm an increased magnetic signal to noise ratio with in plane atomic resolution.

## I. INTRODUCTION

Since its first experimental demonstration [1], electron magnetic chiral dichroism (EMCD) has attracted great interest in physics and materials science because it offers the potential to study the magnetic properties of materials in the transmission electron microscope (TEM) with atomic spatial resolution. [2,3,4] In the earliest formulation of EMCD proposed by Schattschneider et al. [1, 5], the measurement of a dichroic signal is based on the use of a parallel electron beam, a two (or three) beam orientation of the crystalline sample and the recording of electron energy-loss (EEL) spectra at two specific positions in the diffraction plane. The drawbacks to this approach include limited spatial resolution [6] (typically several nm), poor signal-to-noise ratio (SNR) and a strong dependence of the strength of the dichroic signal on sample thickness, with numerical simulations in Ref. [5] reporting that the signal can be close to zero at some sample thicknesses. Whereas the principle of the technique relies on the detection of a change in the orbital angular momentum of the inelastically scattered wave, its standard formulation only makes use of post-selection of the scattered momentum.

A recent advance in electron microscopy has involved the introduction of electron vortex beams [7, 8, 9], including the possibility to create atomic-sized electron vortices [10, 11]. Electron beams that have a given topological charge can in principle be focused onto a single atomic column and induce atomic excitations, with different intensities for transitions in which the magnetic quantum number changes by $\Delta m = \pm 1$. [12] For example, for *2p→3d* transitions in magnetic transition metals differences in energy-resolved diffraction patterns are expected for vortices with opposite orbital angular momentum (OAM) of ±ℏ because of the different populations of spin-up and spin-down *3d* electronic states. [13,14,15] These methods solve some of the issues of the classical EMCD formulation as they can be performed in zone axis conditions thus increasing the achievable spatial resolution. [4] [16] [17].

Unfortunately, this approach is difficult for three reasons. First, it requires two measurements using opposite electron vortex beams (typically with $l = +1$ and $l = -1$), which cannot easily be performed on exactly the same region due to sample and probe drift. Second, an electron vortex does not conserve its OAM while propagating through a crystal, as the free space cylindrical symmetry is broken by the crystal potential [18,19,20]. The probing electron in the crystal is therefore not in a quantum mechanical state with well-defined OAM during its propagation and parts of it acquire different OAM, reducing the intensity of the ±ℏ components and, in turn, the dichroic signal [16]. Third, it is experimentally challenging to prepare a high-quality coherent atomic size electron vortex beam. [10]

These drawbacks can be solved if a standard electron probe can be used and post interaction analysis in terms of OAM can be performed on the inelastically scattered electrons,[21] as proposed in Ref. [22] for amorphous materials. Accordingly, here we propose a new approach for performing EMCD experiments based on post-selection in both energy and OAM of inelastically scattered electrons from a crystalline magnetic sample. The proposed setup is shown schematically in Figure 1.

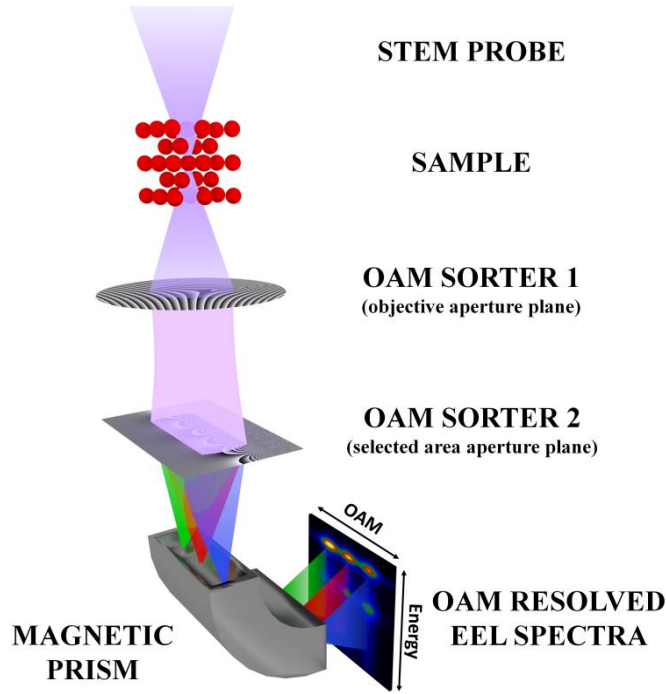

**Figure 1**: Proposed experimental setup needed to perform an OAM-resolved electron energy-loss (EEL) spectroscopy experiment in a scanning TEM (STEM).

A conventional focused electron probe, which has been formed using a circular aperture, is used to image a magnetic sample along an high symmetry zone axis. Two phase elements (referred to as OAM sorters) [23, 24, 25], which are located in the objective and selected area diffraction planes of the microscope column, are used to spatially separate the OAM components of the electron beam. If these elements have an appropriate orientation, then the electrons can be analyzed using an energy-selecting spectrometer (whose aperture is located in the diffraction plane of Sorter 2) to achieve double dispersion of the electrons (in orthogonal directions) in both energy and OAM [26].

In contrast to the use of an electron vortex beam as a probe, this setup allows the two spectra that are needed for an EMCD measurement to be recorded simultaneously, as already reported for some classical EMCD measurements [27,28,29,30,31]. Furthermore, the sample can be oriented along a high symmetry direction, allowing for atomic resolution. In addition, as we explain below, a large fraction of the inelastic signal can be used to record the EMCD spectra, being the only limiting factor the efficiency of the sorting apparatus.

## II. THEORY

We begin by describing the theory that is required to simulate OAM-resolved core loss spectra $I(l, \Delta E)$ in crystalline samples using multislice calculations. Interestingly, the formalism that is required to describe EMCD spectra becomes clearer once a description in terms of electronic OAM is used, when compared to a description in terms of linear momentum.

The combined use of OAM sorters and an EEL spectrometer provides access to the quantities

$$I(l, \Delta E) = Tr[\hat{\rho}_F \hat{P}_{lE}] \quad (1)$$

where $\hat{\rho}_F$ is the electron density matrix after passing the sample, $\hat{P}_{lE} = \int dk |l, k, E><l, k, E|$ and $E = E_0 - \Delta E$ is the energy of an electron that has lost energy $\Delta E$. The quantum states $|l, k, E>$ are characterized by OAM values $\hbar l$ and energies $E$, while $k$ is a quantum number whose physical meaning is explained below. The quantity $I(l, \Delta E)$, which is experimentally available using the proposed setup, defines the number of electrons that have energy $E_0 - \Delta E$ and are found in a state with OAM equal to $\hbar l$. The dichroic signal can be obtained as a difference between $I(+1, \Delta E)$ and $I(-1, \Delta E)$.

We make use of the approach described in Ref. [32] to simulate OAM-resolved core loss spectra. We adapt the formalism, which is reviewed here for completeness, from a momentum-defined to an angular-momentum-defined final state basis, based on the following approximations:

- the electrons are assumed to undergo single inelastic processes while passing through the sample. This approximation is justified by the low probability of core loss excitation;
- the energy of the probing electrons is assumed to be 300 keV. We work in the paraxial approximation, for which the electron wavefunction can be written $\Psi(\mathbf{r}) = e^{ik_z z}\phi(\mathbf{r}^\perp; z)$, where we use the definition $\phi(\mathbf{r}^\perp; z) \equiv <\mathbf{r}^\perp|\phi_z>$ below for the transverse wavefunction at height $z$;
- we use the z-locality approximation [33, 34], where the state of an electron after inelastic scattering from an atom located at $\mathbf{a}$ can be written $|\phi_{z=z_a}^{i,f_a}> = -i\sigma \hat{V}_{i\rightarrow f}^{proj}|\phi_{z=z_a}^{el}>$, the atom at $\mathbf{a}$ experiences a transition $|i>\rightarrow |f>$, $|\phi_{z=z_a}^{el}>$ is the electron state at $z = z_a$ before inelastic scattering, $\hat{V}_{i\rightarrow f}^{proj}$ is defined in Eq. 4 in Ref. [32] and $\sigma$ is a parameter that depends only on the energy of the electron beam;
- we consider a highly symmetrical cubic crystal, i.e., *bcc* iron. This assumption simplifies the definitions of the quantities of interest. However, the conclusions that we obtain for this system can be generalized to less symmetrical samples.

The final density matrix $\hat{\rho}_F$ can be written in single scattering approximation as

$$\hat{\rho}_F = \sum_{\mathbf{a},f,i} |\phi_{z=t}^{i,f_a}><\phi_{z=t}^{i,f_a}|$$

being $|\phi_{z=t}^{i,f_a}>$ the state of an electron which has experienced an inelastic scattering event at $z = z_a$ and has then elastically propagated up to the exit surface of the sample ($z = t$). So starting from Eq. 1, we can write

$$I(l, \Delta E) = \sum_{a,i,f} \int_0^{K_{Max}(\alpha)} k dk |< l, k|\phi_{z=t}^{i,f_a}>|^2 \delta(\Delta E + \varepsilon_f - \varepsilon_i) = I_\alpha(l, \Delta E) \quad (2)$$

The states $|l, k>$ are defined such that $<r^\perp|l, k> = e^{il\varphi} J_{|l|}(k|r^\perp|)$, i.e., they correspond to Bessel beams with orbital angular momentum $l\hbar$ and transverse wavevector $k$. The integral over $k$ in Eq. 2 is performed up to $K_{Max}(\alpha) = 2\pi\alpha/\lambda$, where $\alpha$ is the numerical aperture of the OAM sorter in the back focal plane of the objective lens, while $\lambda$ is the electron de Broglie wavelength, which takes a value of 1.97 pm at 300 kV. This also defines the maximum collection angle of the experimental setup, assuming no limitation on the energy spectrometer. The electron state immediately after the inelastic event at $z = z_a$ (i.e. $|\phi_{z=z_a}^{i,f_a}>$) is related to the one at $z = t$ through the unitary evolution operator $U(t, z_a)$ defined by Eq. 3 of Ref.[32], by the relation $|\phi_{z=t}^{i,f_a}> = U(t, z_a)|\phi_{z=z_a}^{i,f_a}>$: therefore, by direct substitution in Eq. 2 of this relation and exploiting the property $U(z_a, t) = U^{-1}(t, z_a) = U^\dagger(t, z_a)$ we obtain

$$I_\alpha(l, \Delta E) = \sigma^2 \sum_{a,i,f} \int_0^{K_{Max}(\alpha)} k dk < l, k|U^\dagger(z_a, t) \hat{V}_{i \to f}^{proj}|\phi_{z=z_a}^{el}><\phi_{z=z_a}^{el}|\hat{V}_{i \to f}^{proj} U(z_a, t)|l, k$$
$$> \delta(\Delta E + \varepsilon_f - \varepsilon_i)$$

Now we recognize that $U(z_a, t)|l, k> = |lk^{BP}>$, i.e. it corresponds to the wavefunction obtained elastically back-propagating the Bessel beam $|l, k>$ from the bottom of the sample to $z = z_a$: therefore it is immediate to realize that

$$I_\alpha(l, \Delta E) = \sigma^2 \sum_{a,i,f} \int_0^{K_{Max}(\alpha)} k dk |< lk^{BP}|\hat{V}_{i \to f}^{proj}|\phi_{z=z_a}^{el}>|^2 \delta(\Delta E + \varepsilon_f - \varepsilon_i) =$$
$$\int_0^{K_{Max}} I(l, k, \Delta E) k \, dk$$

with

$$I(l, k, \Delta E) = \sigma^2 \sum_{a,i,f} |< lk^{BP}|\hat{V}_{i \to f}^{proj}|\phi_{z=z_a}^{el}>|^2 \delta(\Delta E + \varepsilon_f - \varepsilon_i) \quad (3)$$

the function which describes the dependence of the OAM resolved loss function from the transverse scattering wavevector $k$.

Proceeding as in Ref. [32], it is possible to rewrite $I(l, k, \Delta E)$ as

$$I(l, k, \Delta E) = \sigma^2 \sum_a \int d\mathbf{k}_1 \dots \int d\mathbf{k}_4 \, D^*(\mathbf{k}_1; l, k) C(\mathbf{k}_2) D(\mathbf{k}_3; l, k) C^*(\mathbf{k}_4) \frac{S_a(\tilde{q}, \tilde{q}', \Delta E)}{\tilde{q}^2 \tilde{q}'^2} e^{i(q-q')a}, \quad (4)$$

where the sum over $a$ is performed over atomic positions at which the inelastic process of interest can occur. Given the infinitesimal distribution in $k$, $D(\mathbf{k}_1; l, k)$ is the three-dimensional Fourier transform of an elastically scattered Bessel beam with topological charge $l$, transverse wavevector $k$ and energy $E_0 - \Delta E$ propagating from the exit to the entrance surface of the crystal.

In this expression, the energy dependence of $D(\mathbf{k}_1; l, k)$ is neglected, as we focus on a small range of energy losses $\Delta E$ in the interval $[\Delta E_{min}, \Delta E_{max}] = [690, 750]$eV. The propagation of an electron of energy $E_0 - \Delta E_{min}$ is similar to that of one with energy $E_0 - \Delta E_{max}$ because $E_0 \gg \Delta E_{max}, \Delta E_{min}$.

Analogously, $C(\mathbf{k}_2)$ is the three-dimensional Fourier transform of an incident beam that propagates elastically in the material with energy $E_0$, from the bottom to the top of the sample. Both $D(\mathbf{k}_1; l, k)$ and $C(k_2)$ can be obtained by performing three-dimensional Fourier transforms of the wavefunction inside the crystal, computed using a multislice calculation.
The quantities $\tilde{q}$, $\tilde{q}'$, $q$ and $q'$ appearing in Eq. 4 are defined as [32]

$$\tilde{q} = k_1^\perp - k_2^\perp + q_{\Delta E}\hat{z} \qquad \tilde{q}' = k_3^\perp - k_4^\perp + q_{\Delta E}\hat{z}$$

$$q = k_1 - k_2 \qquad q' = k_3 - k_4$$

$$q_{\Delta E} = k_z^f - k_z^i$$

$q$ and $q'$ are the elastic scattering wavevectors in the crystal, while $\tilde{q}$ and $\tilde{q}'$ differ from the precedent ones only through the $z$ components, given in this case by $q_{\Delta E}$, which corresponds to the difference between the final and the initial $z$-components of the electron beam's wavevector; as detailed in Ref. [32], $\tilde{q}$ and $\tilde{q}'$ directly appear in the expression of the loss function once the $z$-locality approximation is taken into account.

In Eq. 4, $S_a(\tilde{q}, \tilde{q}', \Delta E)$ is the atomic mixed dynamic form factor [35], which provides the energy dependence of $I(l, k, \Delta E)$: further it also plays a crucial role in the OAM dependence (which can be seen by expanding it in spherical harmonics), in the k dependence (by coupling different k components in the incident and outgoing beams), and in the thickness dependence of the detected signal. Working within the dipolar approximation, we can write [36]

$$S_a(\tilde{q}, \tilde{q}', \Delta E) = \tilde{q}\mathbb{N}_a(\Delta E)\tilde{q}' + i\boldsymbol{\mathcal{M}}_a(\Delta E)[\tilde{q} \times \tilde{q}'],$$

where $\boldsymbol{\mathcal{M}}_a(\Delta E)$ is a vector that describes the magnetic properties of the sample, while $\mathbb{N}_a(\Delta E)$ is a real symmetric tensor, taking into account the non-magnetic contributions to the signal. We assume that the magnetic field of the objective lens is sufficiently strong to saturate the magnetization in the sample along $z$: only the $z$ component of $\boldsymbol{\mathcal{M}}_a(\Delta E)$ is then non-zero. For a cubic crystal whose axes are parallel to $\hat{x}$, $\hat{y}$ and $\hat{z}$, $\mathbb{N}_a(\Delta E)$ becomes a diagonal tensor, with all of the diagonal elements equal to each other. [37]. A set of functions [38] can be introduced allowing to group together the different terms that are associated with dynamical diffraction effects. We define

$$Q_a^i(l, k) = \sigma \int dk_1 \int dk_2\, D^*(k_1; l, k) C(k_2) \frac{\tilde{q}_i}{\tilde{q}^2} e^{i(q-q')a} \qquad i = x, y, z \quad (5)$$

$$X_i(l, k) = \sum_a |Q_a^i(l, k)|^2 \quad (6)$$

$$\tilde{S}(l, k) = -2 Im\left[\sum_a Q_a^x(l, k) Q_a^y(l, k)^*\right] \quad (7)$$

The loss function then assumes the form:

$$I(l, k, \Delta E) = \sum_{i=x,y,z} \mathbb{N}_i(\Delta E) X_i(l, k) + \boldsymbol{\mathcal{M}}(\Delta E)\tilde{S}(l, k) \quad (8)$$

where we have neglected the dependence of $\mathbb{N}$ and $\boldsymbol{\mathcal{M}}$ on atom type, as we focus on energy losses produced by atoms of the same type and the same coordination geometry in the crystal.
The functions $X_i(l, k)$ and $\tilde{S}(l, k)$ describe the effect of dynamical diffraction in the crystal and need to be tuned to achieve optimal experimental conditions; further, as shown in the Appendix A, the functions $X_i$ and $\tilde{S}(l, k)$ satisfy the properties:
$$X_i(l, k) = X_i(-l, k) \text{ and } \tilde{S}(l, k) = -\tilde{S}(-l, k).$$

Exploiting these relations, we define a relative dichroic function $R_{|l|}(k)$ for cubic crystals (the ones considered in the present work), for which $N_i(\Delta E) = N(\Delta E)$ for $i = x, y, z$.

$$100 \frac{I(l,k,\Delta E) - I(-l,k,\Delta E)}{I(l,k,\Delta E) + I(-l,k,\Delta E)} = \frac{\mathcal{M}(\Delta E)}{N(\Delta E)} R_{|l|}(k)$$

In practice $R_{|l|}(k)$ describes how the dynamical effects associated to the electron beam elastic propagation inside the crystal affect the strength of the dichroic signal (i.e. the difference between $I(l,k,\Delta E)$ and $I(-l,k,\Delta E)$ normalized by its sum) evaluated at a fixed transverse wavevector $k$: even if presented for the case of cubic materials, such a definition could be also generalized to crystals with other symmetries, even if such a defined separation among the energy dependence (here represented by the ratio $\frac{\mathcal{M}(\Delta E)}{N(\Delta E)}$) and the dynamical diffraction effects would not be so direct. During the proposed experiment we measure the integral of these functions over a range of vectors determined by the semi-collection angle $\alpha$ of the OAM spectrometer, i.e. $I_\alpha(l,\Delta E)$.
Therefore we define integrated OAM dependent dynamical coefficients $\chi_i(l,\alpha)$ and $\Lambda(l,\alpha)$

$$\chi_i(l,\alpha) = \int_0^{K_{Max}(\alpha)} k dk X_i(l,k)$$

$$\Lambda(l,\alpha) = \int_0^{K_{Max}(\alpha)} k dk \tilde{S}(l,k)$$

which respectively correspond to the non magnetic and magnetic integrated contributions to the OAM loss function at a certain $l$: the dependence from $\alpha$ comes from the fact that their values change by modifying the experimental semi-collection angle. Notice that these quantities satisfy the same relations valid for their not-integrated counterparts, i.e. $\chi_i(l,\alpha) = \chi_i(-l,\alpha)$ and $\Lambda(l,\alpha) = -\Lambda(-l,\alpha)$.
Using these definitions and considering Eq.8 we can now express the integrated OAM resolved loss function as

$$I_\alpha(l,\Delta E) = \sum_{i=x,y,z} N_i(\Delta E)\chi_i(l,\alpha) + \mathcal{M}(\Delta E)\Lambda(l,\alpha) \quad (9)$$

We can now exploit these integrated quantities to define a dichroism function $D^l{}_\alpha(\Delta E)$ for a certain OAM $l$, which can be determined directly from the experimentally measured OAM loss function accessible from the experimental set up outlined in the precedent section: such a function is given by

$$D^l{}_\alpha(\Delta E) = 100 \frac{I_\alpha(l,\Delta E) - I_\alpha(-l,\Delta E)}{\max_{\Delta E}(I_\alpha(l,\Delta E) + I_\alpha(-l,\Delta E))} \quad (10)$$

The dichroism function is normalized to the maximum combined intensity of the $+l$ and $-l$ loss functions over the entire energy range in order to leave the energy dependence of $D^l{}_\alpha$ only being determined by the difference between $I_\alpha(l,\Delta E)$ and $I_\alpha(-l,\Delta E)$. This normalization is convenient, because one can directly apply EMCD sum rules [39,40] to such dichroic function.
Now, using Eq. 9, the properties of $\chi_i(l,\alpha)$ and $\Lambda(l,\alpha)$ and assuming a cubic crystal it is simple to demonstrate that:

$$D^l{}_\alpha(\Delta E) = 100 \frac{\mathcal{M}(\Delta E)\Lambda(|l|,\alpha)}{\max_{\Delta E}(N(\Delta E)\sum_{i=x,y,z}\chi_i(|l|,\alpha))} \quad (11),$$

This function describes the strength of the dichroic signal for a certain $|l|$, which can be obtained experimentally and also its dependence from the energy loss and the collection semi-angle chosen for the experiment.
All these quantities have been computed using a modified version of the software MATSv2 [38]: previous releases of the software were based on a mixed multislice / Bloch wave approach for the propagation of the incoming and reciprocal beam respectively. In order to allow for the back propagation of the reciprocal Bessel beam we updated the code to a full multislice for both direct and reciprocal wave. As a consequence the number of terms of the summation in eq. 5 is vastly increased, requiring a more elaborate way of grouping them (see ref. 38) to speed up the calculation.

## III. RESULTS AND DISCUSSION

In this section, we present calculations of the non-magnetic signal $X_i(l,k)$ and the magnetic signal $\tilde{S}(l,k)$ for $l=[-2;2]$ and different values of $k$ for *bcc* Fe oriented along the [001] direction. In the simulations, we use an orthogonal supercell of size $[20\times20]a$ (where $a=0.287$ nm is the lattice parameter of Fe), considering thicknesses of 40 nm and 20 nm. The electron probe is centered on an Fe column at (0,0) and the signals $X_i(l,k)$ and $\tilde{S}(l,k)$ are evaluated by summing the products $D^*(\mathbf{k}_1;l,k)C(\mathbf{k}_2)$ for a convergence parameter of $5\cdot10^{-9}$.

Figure 2a shows the magnetic term $\tilde{S}(l,k)$ for different values of $l$, taking a STEM probe with semi-convergence angle of 7.3 mrad and considering a 40 nm thick sample. In agreement with the conclusions reported in the Appendix A (i.e., $\tilde{S}(l,k)=-\tilde{S}(-l,k)$), for $l=+1$ this function is equal in modulus but opposite in sign to that evaluated for the opposite topological charge and is zero if the orbital angular momentum is taken to be zero. It is also negligible for larger values of $|l|$, as we are working in dipolar approximation, which is a reasonable assumption as, in the angular range considered in this work, dipole-allowed transitions are by far dominant [41].

Figure 2b shows the non-magnetic term $\sum_i X_i(l,k)$ (using the same probe and the same thickness as in Figure 2a), which is non-zero for $l=0$. It decreases more rapidly than the non magnetic contribution for $l=\pm1$ and is peaked at smaller scattering angles ($\lambda k$). In contrast to the magnetic terms, these quantities are independent of the sign of the OAM (as $\sum_i X_i(1,k)=\sum_i X_i(-1,k)$), while they become negligible for $|l|>2$. The presence of intensity for $|l|>1$ is only dictated by the fact that OAM is not conserved during elastic scattering in the crystal, thus giving rise to signals for larger $|l|$.

Notice that both the magnetic and non magnetic contributions for $|l|=1$ are peaked at a scattering angle of about 7 mrad: in order to test if the position of this maximum is dependent from the probe convergence and/or the sample thickness, we have performed calculations for a sample of 20 nm, with various STEM probes: the results for the magnetic term evaluated for $l=-1$ are presented in Figure 2c. It is simple to realize that these functions are all peaked at 7 mrad, independently from both the sample thickness and the convergence of the STEM probe.

Finally in Figure 2d we display the relative dichroic function $R_{|l|=1}(k)$ in terms of the transverse scattering angle $\lambda k$ for a STEM probe of 7.3 mrad and samples of 20 and 40 nm.

As described in the precedent section, this function describes the effect of the electron dynamical diffraction in the crystal on the strength of the dichroism, as a function of $k$: the fact that the overall trends of these functions do not strongly change modifying the sample thickness practically suggests that the intensity of the dichroic signal within this experimental set up is expected to be only weakly dependent from the thickness itself.

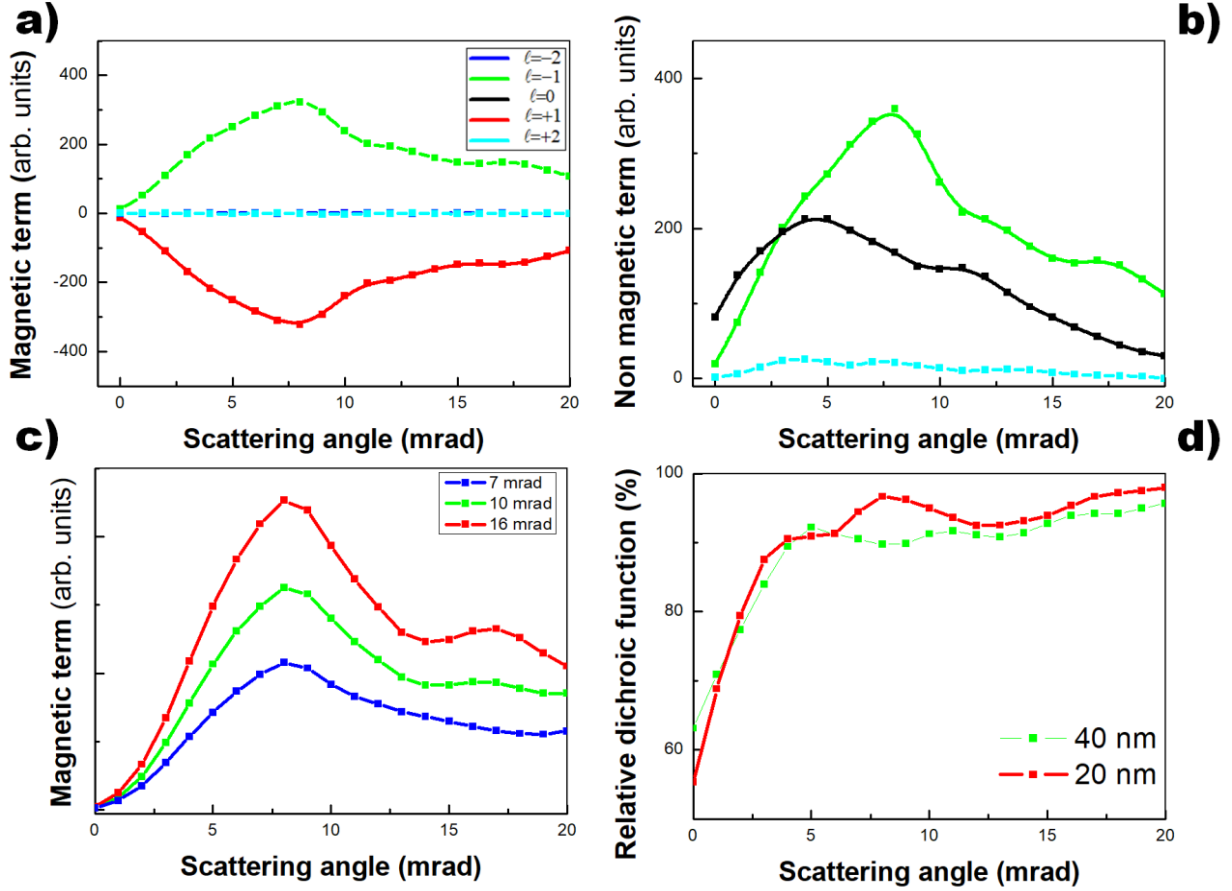

**Figure 2**: a) Magnetic $\tilde{S}(l,k)$ and b) non-magnetic $\sum_i X_i(l,k)$ terms plotted *vs* scattering angle $\lambda k$ for $l$ in the interval [-2;+2], assuming a STEM probe of 7.3 mrad and a bcc Fe sample of 40 nm. c) Functions $\tilde{S}(l,k)$ for $l = +1$ computed for a Bcc Fe sample with thickness of 20 nm for different STEM convergence of the STEM probe: notice that as in the case of the sample of 40 nm this function is still peaked at about 7 mrad and this also appears to be independent from the convergence of the probe itself. d) Relative dichroic function $R_{|l|=1}(k)$ plotted *vs* scattering angle computed for a samples of 20 and 40 nm, taking an incident beam with semi-convergence angle of 7.3 mrad: notice that the overall trend of this function seems to be almost independent from the sample thickness.

This analysis points out that with semi-collection angles of the order of 7-10 mrad, an intense dichroic signal is expected to be observed. It is clear that a finer optimization of the experimental conditions for OAM-resolved EMCD will require to take into account the real efficiency of the OAM sorters, which will be the object of future experimental work.

In the following we will assume to work with semi-collection angles of 7 mrad, a range accessible with the OAM spectrometer already presented in Ref. [23].

It is interesting to find a possible rationale for the dependence of $\tilde{S}(\pm 1, k)$ and $X_i(\pm 1, k)$ from the transverse wavevector $k$: in particular the reason for which these functions are peaked at the same value of $k$, and they decrease in strength by increasing $k$, independently from the probe convergence and the sample thickness.

In Appendix B we provide an approximate evaluation of the function $Q_a^i(l,k)$ for i = x,y, as the imaginary part of the product of these two function defines $\tilde{S}$ (see Eq. 7) and their square modulus gives important contribution to the non-magnetic dynamical coefficient (see Eq. 6). What emerges is that the larger is, on average, the strength of the OAM component $m = \pm 1$ of the back propagating Bessel beam $|l,k>$, the larger will be the value of $Q_a^{x,y}(l,k)$ computed for that specific pair of indices $(l,k)$. So taking $l = \pm 1$, the approximate reasoning exposed in Appendix B suggests us that $\tilde{S}(\pm 1, k)$ and $X_i(\pm 1, k)$ are peaked at about 7 mrad because Bessel beams $|l = \pm 1, \lambda k =$ 7 mrad > have on average along $z$ a larger $m = \pm 1$ contribution with respect to those associated with larger transverse wavevectors. To confirm this point we performed multislice calculation [42,43] of the propagation of Bessel beams with various wavevectors $k$ and $l = +1$ in a bcc Fe sample with thickness of 40 nm and we computed the OAM decomposition for the beam at different depths $z$: in other words, the coefficients $D_m^{l=+1,k}(z)$ defined in Appendix B have been evaluated for various $m$.

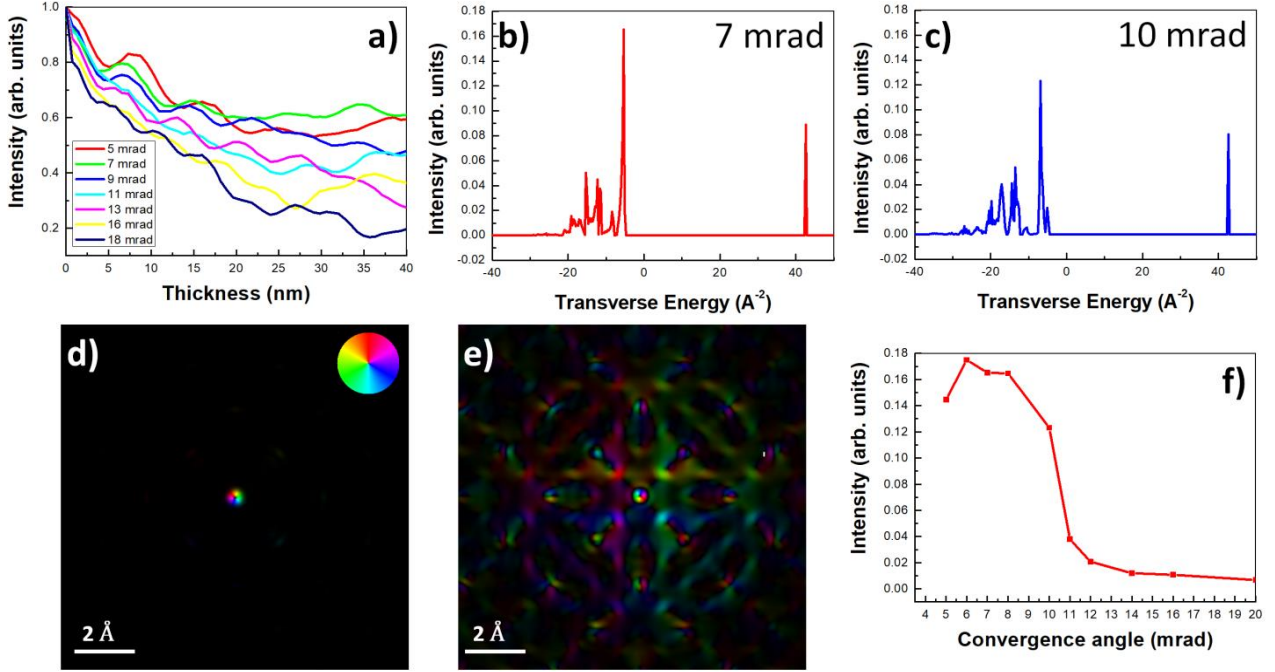

**Figure 3** a) Functions $D_{m=1}^{l=1,k}$ computed for various transverse wavevectors at different $z$ in the crystal: notice that for large $k$ the component m = 1 decreases more rapidly than for k close to 7 mrad. b) and c) Spectrum of the Bloch wave excited by a Bessel beam with transverse wavevector of 7 mrad and 10 mrad respectively. d) the complex wavefunctions corresponding to the states having 42 and -5Å$^{-2}$ transverse energy (modulus and phase encoded in brightness and color). f) excitation of the state presented in d) as a function of the transverse wavevectors: notice that the excitation is maximum around 6-7 mrad, while it rapidly decreases for larger values of $k$: as such a state is characterized by a strong $l = 1$ component it is reasonable to expect the Bessel beams with $k$ in this range to better preserve its original OAM.

These coefficients, for some transverse wavevectors are shown in Figure 3a) (see legend): we see that for $\lambda k$ slightly smaller than 7 mrad $D_m^{l=+1,k}(z)$ is (on average along $z$) smaller than the one computed for $\lambda k = 7$ mrad for every thickness; at the same time, by increasing $\lambda k$ beyond 7 mrad we observe that these coefficients progressively decrease in intensity as functions of $z$. This conclusions are in qualitative agreement with what has been deduced from the approximate analytical treatment exposed in Appendix B.

To further clarify the reason why a Bessel vortex beam $|l = \pm 1, \lambda k = 7 \text{ mrad} >$ is able to better preserve its OAM with respect to Bessel beams with larger or smaller transverse wavevectors we have performed Bloch wave calculations using our custom software B_WISE following the procedure reported in [44]: the incident probe is sampled, in Fourier space, into many points and for each one of these points we performed a Bloch wave calculation following the original algorithm proposed by Metherell. The resulting Bloch coefficients are then summed together, taking into account the appropriate phase, in narrow ranges of the transverse kinetic energy (or anpassung parameter) to produce a spectral representation of the convergent probe propagating inside the crystal.

In figure 3 b) and c) we present the transverse energy spectra of the excited Bloch states intensities for $\lambda k = 7$ mrad and $\lambda k = 10$ mrad Bessel beams respectively. Two main peaks appear in the spectra: the first one is located at an energy of approximately 42 $\dot{A}^{-2}$ and correspond to highest energy Bloch state, usually called 1s state because of its resemblance with the homonymous atomic state, and the second one located at about -5 $\dot{A}^{-2}$. The wavefunctions of the two states are reported in figure 3 d) and 3 e) respectively. Both the states show a well-defined $l = 1$ vortex along the central atomic column, whereas other lower energy states (not shown here) are delocalized over many unit cells, therefore it is easy to assume that these two states are responsible for the OAM conservation. Noticeable, the excitation of the -5 $\dot{A}^{-2}$ state is strongly affected by the Bessel beam's transverse wavevector and drastically decreases passing from $\lambda k = 7$ mrad to $\lambda k = 10$ mrad. The excitation of the 42 $\dot{A}^{-2}$ state is instead less affected. What naturally follows is to trace the behaviour of the excitation of the -5 $\dot{A}^{-2}$ state over a wide range of angles, as reported in figure 3f). This shows a maximum at 6-7 mrad and overall shape strictly follows the k dependence of the cross sections reported in figure 2).

It is important to understand if the approach described here provides access to magnetic properties with atomic spatial resolution. According to the definition of $\tilde{S}(l, k)$ in Eq. 7, the sum is performed over all of the magnetic atoms in the sample, not only those in the column on which the STEM probe is focused. In order to clarify this point, we evaluated the contributions to the magnetic signal (Figs 4a and 4c) and the non-magnetic signal (Fig. 4b and 4d) for $l = +1$ for both the atoms in the column on which the probe is centered and those in the neighbouring columns [38, 45]. The magnitudes of the terms are represented as solid spheres centered on each atomic position. The radius of each sphere is directly proportional to the modulus of that term. The same information is encoded in the spheres' colors on a logarithmic scale (see color bars in Fig. 4). The contributions decrease rapidly for atoms that are not on the column centered at the origin, suggesting that the measured signal comes almost entirely from the atomic column of interest. Atomic spatial resolution in measuring the magnetic properties of the material is therefore expected.

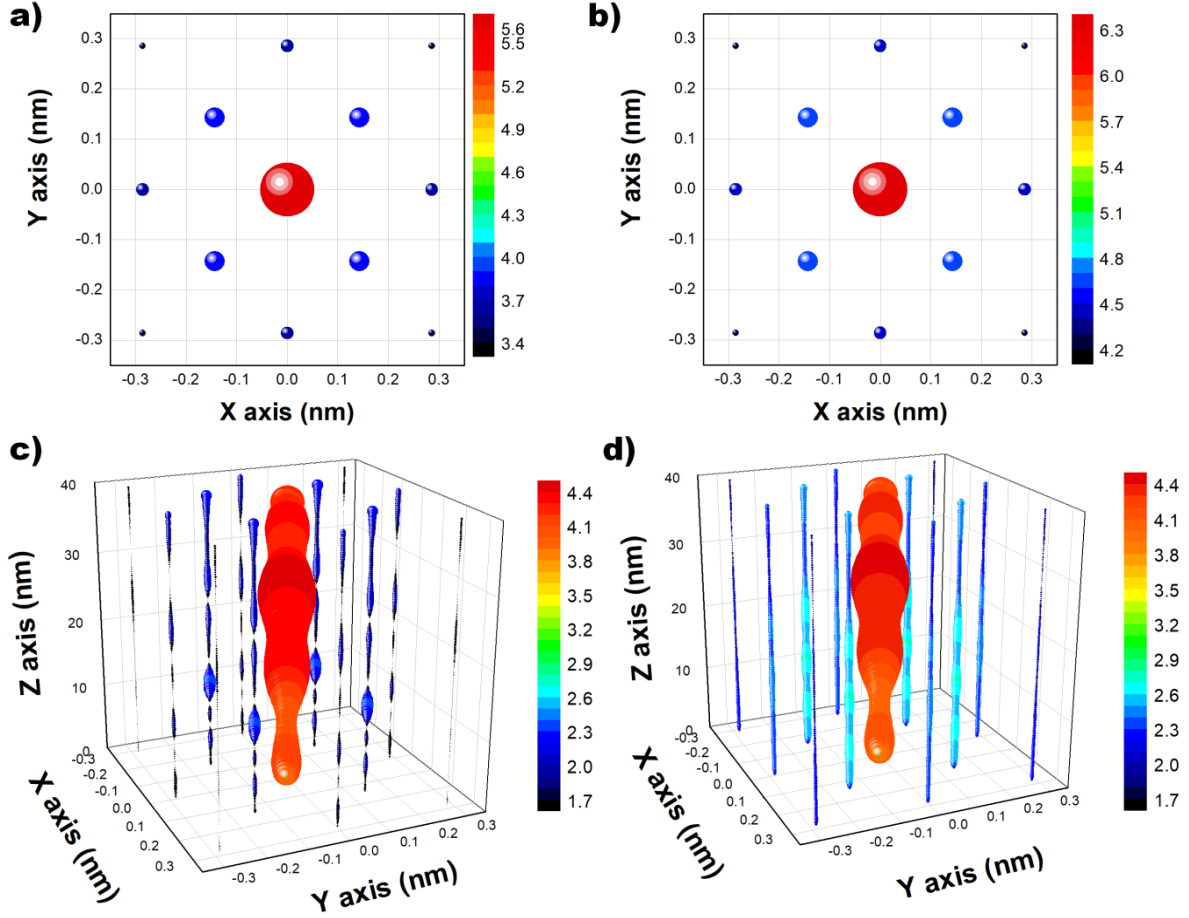

**Figure 4**: a) and c) Two-dimensional and three-dimensional views, respectively, of contributions to the function $\Lambda(l = 1, \alpha = 7 \text{ mrad})$ from atoms close to and on the atomic column on which the electron probe is centered. Each atom is surrounded by a sphere, whose radius (and color, on a logarithmic scale) is proportional to the strength of the contribution from that atom. On increasing the distance from the atomic column at $(0, 0)$, there is a rapid decrease in the contribution from the atoms to $S$. b) and d) Corresponding depictions for the non-magnetic part of the signal.

Once the functions that describe dynamical diffraction of the electron beam in the sample are known, it is possible to estimate OAM-resolved EEL spectra using Eq. 9. The functions $N(\Delta E)$ and $\mathcal{M}(\Delta E)$ have been evaluated using first principles calculations [5, 45], while the maximum collection angle was fixed to 7 mrad. First principles calculations were performed for *bcc* Fe using the WIEN2k package [46] using the generalized gradient approximation for the exchange-correlation functional [47] and including spin-orbit coupling effects. The atomic sphere radius of Fe was set to 2.33 Bohr radii, the basis size cut-off was $RK_{max} = 8.0$ and Brillouin zone integrations were performed using a modified tetrahedron method with 10000 **k** points. The upper panel of Fig. 4a shows the resulting spectra for different values of OAM, while lower panel shows the dichroic function defined in Eq. 10. The quantity $D^{|l|=1}{}_\alpha(\Delta E)$ (computed for a semi-collection angle $\alpha = 7$ mrad) reaches values of ~15% (or more) at both the $L_2$ and the $L_3$ Fe edges. This quantity is much larger than the relative dichroic signal that is observed using conventional approaches to EMCD.

As outlined above, this approach should provide access to double dispersion, in perpendicular directions, as a function of energy and orbital angular momentum. However, in reality the spectra shown in Fig. 5a are different from those that we expect to measure. In order to obtain results that are similar to those expected from a real life experiment, our treatment must include broadening in energy of the electron beam (as it is not perfectly monochromatic) and finite energy resolution of the OAM sorters. What we expect to observe in practice is

$$\Gamma(l, \Delta E) = I_{\alpha=7\text{mrad}}(l, \Delta E) \otimes f(l, \Delta E) \quad (12)$$

where $f(l, \Delta E)$ is the product of two Gaussian functions describing broadening introduced by the experimental setup. The OAM is treated as a continuous variable, since the sorter can transform a vortex with topological charge $l$ in a spot centered at coordinate $Cl$ [23, 24, 25] with a lateral extension equal to $C\Delta l$ (where $C$ is a constant that depends on the sorter parameters and is assigned a value of unity for simplicity in the images presented in the main text).

Figure 5b shows the result of this convolution procedure for $\Delta l = 0.5\hbar$ and $\Delta E = 0.7 eV$, as this is the resolution expected using a sorter in a fan-out configuration, as recently demonstrated for optical sorters [48]. Despite the finite OAM resolution, which introduces partial mixing of the signal at $l = 0$ with that for $l = \pm 1$, strong asymmetry between $\Gamma(+1, \Delta E)$ and $\Gamma(-1, \Delta E)$ is observable.

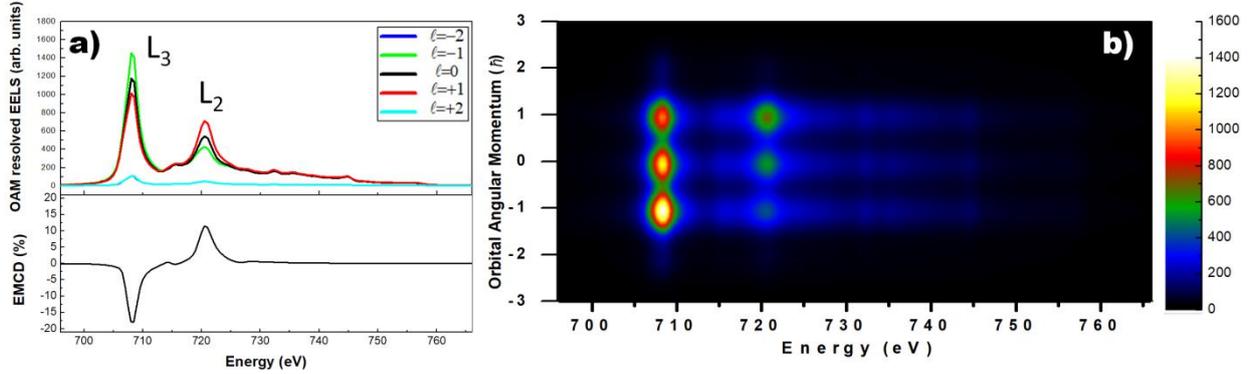

**Figure 5:** a) Upper panel: OAM-resolved EEL spectra computed for $l$ in the range [-2;2] for a collection semi-angle of 7 mrad and sample thickness of 40 nm. Lower panel: $D^{|l|=1}(\Delta E)$ defined in Eq. 10, in which the dichroism strength reaches values of ~18% for both the $L_2$ and the $L_3$ edge of Fe. b) Convolution of the spectra shown in a) with the product of two Gaussian functions describing broadening in OAM (0.5$\hbar$) and energy (0.7 eV) introduced by the OAM sorters and non-monochromaticity of the electron beam.

It should be remarked here that the inelastic signal that we observe for $l = \pm 2$ is not due to electron transitions with a change in OAM equal to $\pm 2\hbar$, as our approach is based on a dipolar approximation, but it is only due to a lack of OAM conservation for a beam that propagates in a crystal. More precisely, once an electron has experienced an inelastic event, it keeps propagating in the crystal but its OAM is not conserved and acquires components corresponding to $l \neq 0, \pm 1$ (e.g., for $l = \pm 2$), giving rise to a non-zero signal.

Figure 6 shows in-plane spatial mapping of the dichroic function $D^{|l|=1}{}_\alpha(\Delta E)$ obtained by computing OAM-resolved EEL spectra following the procedure outlined above, but scanning the STEM probe across the sample. In Figs 5a and 5b, the function is computed for the $L_2$ and $L_3$ edges, respectively, at energies of $\Delta E = 720$ and $708\ eV$. In absolute value, the strength of the relative dichroism is maximum when the electron beam is centered on an atomic column for both edges, whereas it decreases by a factor of two when the probe is moved between columns, thereby providing information about the magnetic properties of the sample with atomic spatial resolution.

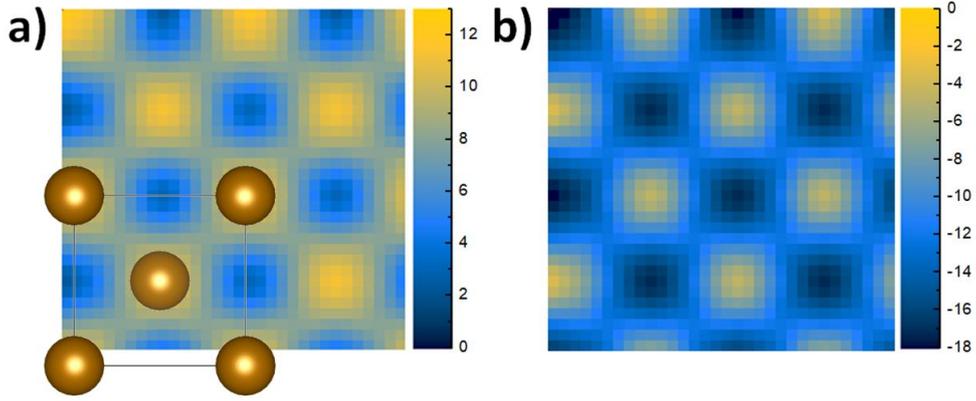

**Figure 6:** Spatial mapping of the function $D^{[l]=1}_{\alpha=7mrad}$ computed for the Fe a) $L_2$ and b) $L_3$ edges. The contrast in the images provides access to atomically resolved mapping of magnetic properties of the sample.

## IV. CONCLUSIONS

We report a new approach that can be used to probe electron magnetic chiral dichroism in the transmission electron microscope by exploiting the recently demonstrated capability of evaluating the orbital angular momentum spectrum of an electron beam. By focusing on the simple case of *bcc* Fe, we introduce the required theoretical framework and we show that this approach should provide strong dichroic signals with atomic spatial resolution, even without requiring very high resolution in orbital angular momentum.

**ACKNOWLEDGEMENTS**

This work is supported by Q-SORT, a project funded by the European Union's Horizon 2020 Research and Innovation Program under grant agreement No. 766970.

## Appendix A

In this appendix, we demonstrate the following properties:

- $S(l, k) = -S(-l, k)$;
- $X_i(l, k) = X_i(-l, k)$

Assuming that the crystalline potential satisfies the symmetry operation

$$V(x, y, z) = V(-x, y, z) \qquad (A1)$$

We begin by explicitly writing the function $D(\boldsymbol{k}_1; l, k)$ in the form

$$D(\boldsymbol{k}_1; l, k)^* = \int dz\, e^{izk_1^z} \int d\boldsymbol{r}_\perp e^{i\boldsymbol{k}_1^\perp \boldsymbol{r}_\perp} \left[U^+(\boldsymbol{r}_\perp, z) e^{il\varphi} J_{|l|}(k|\boldsymbol{r}^\perp|) e^{-izk_f}\right]^* \quad (A2)$$

where, $\boldsymbol{r}_\perp = (x, y)$, $U(\boldsymbol{r}_\perp, z)$ is an evolution operator defined in Ref. [32] and $k_f$ is the projection along the $z$ axis of the wave vector of the inelastically scattered electron. The operator $U(\boldsymbol{r}_\perp, z)$ is invariant under symmetry operations of the crystal, so in our case

$$U(x, y, z) = U(-x, y, z) \qquad (A3)$$

By performing the substitution $x \to -x'$ in Eq. A2, exploiting Eq. A3 and remembering that

$$\int_{-\infty}^{+\infty} dx \to \int_{+\infty}^{-\infty} -dx' = \int_{-\infty}^{+\infty} dx'$$

$$\varphi = \operatorname{atan}\frac{y}{x} \to \varphi' = \operatorname{atan}\frac{y}{-x'} = -\varphi$$

$$J_{|l|}(k|\boldsymbol{r}^\perp|) = J_{|-l|}(k|\boldsymbol{r}^\perp|) \,,$$

we find that

$$D(\boldsymbol{k}_1; l, k)^* = \int dz\, e^{izk_1^z} \int dx' dy\, e^{-ik_1^x x'} e^{ik_1^y y} \left[U^+(-x', y, z) e^{-il\varphi} J_{|l|}(k|\boldsymbol{r}^\perp|) e^{-izk_f}\right]^* = D(-k_1^x, k_1^y, k_1^z; -l, k)^* \,.$$

Using this relation in the definition of $Q_a^x(l, k)$ results in the expression

$$Q_a^x(l, k) = \int d\boldsymbol{k}_1 \int d\boldsymbol{k}_2 D(-k_1^x, k_1^y, k_1^z; -l, k)^* \, C(\boldsymbol{k}_2) \frac{k_1^x - k_2^x}{\tilde{q}^2} e^{iqa} \,, \qquad (A4)$$

which (with the substitutions $-k_1^x \to k'^x_1$ and $-k_2^x \to k'^x_2$) can be rewritten

$$Q_a^x(l,k) = -\int d\mathbf{k'}_1 \int d\mathbf{k'}_2 D(k'^x_1, k^y_1, k^z_1; -l, k)^* C(-k'^x_2, k^y_2, k^z_2) \quad \text{(A5)}$$
$$\times \frac{k'^x_1 - k'^x_2}{\tilde{q}^2} e^{-i(k'^x_1 - k'^x_2)a_x} e^{i(k^y_1 - k^y_2)a_y} e^{i(k^z_1 - k^z_2)a_z}$$

from which it is apparent that $Q^x_{a_x a_y a_z}(l,k) = -Q^x_{-a_x a_y a_z}(-l,k)$.

By proceeding in the same manner, it is possible to show that $Q^y_{a_x a_y a_z}(l,k) = Q^y_{-a_x a_y a_z}(-l,k)$ and $Q^z_{a_x a_y a_z}(l,k) = Q^z_{-a_x a_y a_z}(-l,k)$. Therefore, using the definition of $S(l,k)$, we have

$$S(l,k) = -2Im\left[\sum_a Q^x_{a_x a_y a_z}(l,k) Q^y_{a_x a_y a_z}(l,k)^*\right] = 2Im\left[\sum_a Q^x_{-a_x a_y a_z}(-l,k) Q^y_{-a_x a_y a_z}(-l,k)^*\right] = -S(-l,k) ,$$

where we have exploited the fact that, because of Eq. A1, if we have an atom at position $(a_x a_y a_z)$ then we will have an analogous atom at position $(-a_x a_y a_z)$. By the same reasoning, $X_i(l,k) = X_i(-l,k)$ for $i = x, y, z$.

**Appendix B**

The magnetic and non magnetic parts of the signal, are defined by Eq. 6 and 7, as products of the $Q_a^x(l,k)$ and $Q_a^y(l,k)$ quantities summed over all the magnetic atoms of the sample. In this section we present an approximate analytical calculation of $Q_a^{x,y}(l,k)$ in order to understand the trend of the scattering cross section as a function of the transverse wavevector $k$: in particular our target is to understand the origin of the peak at about 7-8 mrad and the reason for which this function decreases by increasing $k$.

As a first step we assume to consider only the atoms $\mathbf{a}$ located on the column at *(0,0)*: as shown in Section III, this is reasonable as the inelastic signal mainly comes from these atoms. Therefore (considering $Q_a^x(l,k)$, even if a similar treatment is also valid for the *y* component )

$$Q_a^x(l,k) = \sigma \int dk_1^z \int dk_2^z e^{i(k_1^z - k_2^z)a} D^*(\mathbf{k}_1; l, k) C(\mathbf{k}_2) \frac{k_1^x - k_2^x}{|\mathbf{k}_1^\perp - \mathbf{k}_2^\perp|^2 + q_{\Delta E}^2}$$

Calling $\varphi_{inc}(\mathbf{r})$ and $\varphi_{BP}^{lk}(\mathbf{r})$ respectively the incident wavefunction and the back propagated Bessel beam with indices $(l,k)$ in a point $\mathbf{r}$ in the crystal, we develop them in terms of Bessel functions with different topological charges and transverse wavevectors, i.e.

$$\varphi_{BP}^{lk}(\mathbf{r}) = \sum_{k',n} d_n^{lk}(k';z) J_n(k'|r^\perp|) e^{in\phi}$$

$$\varphi_{inc}(\mathbf{r}) = \sum_{k'',m} c_m(k'';z) J_m(k''|r^\perp|) e^{im\phi}$$

The functions $d_n^{lk}(k';z)$ and $c_m(k'';z)$ respectively determine the weight of the corresponding Bessel beam to the functions $\varphi_{BP}^{lk}(\mathbf{r})$ and $\varphi_{inc}(\mathbf{r})$ at a certain depth $z$ in the crystal: the integral over the transverse wavevectors of the modulus square of these terms can be numerically evaluated through multislice calculations and they basically indicate the strength of the component with OAM $n\hbar$ and $m\hbar$ to the overall beam at this depth: clearly, the larger in modulus $d_n(k';z)$ the larger will be the component with OAM $n\hbar$.

Exploiting the Fourier transform of a Bessel beam in two dimensions, i.e.

$$\int d\boldsymbol{r}^\perp e^{-i\boldsymbol{k}^\perp \boldsymbol{r}^\perp} J_l(k|\boldsymbol{r}^\perp|)e^{il\phi} = \delta(|\boldsymbol{k}^\perp|-k)\,e^{il\phi_k}$$

we obtain

$$D^*(\boldsymbol{k}_1;l,k) = \sum_{k',n} \int dz e^{ik_1^z z} d_n^{*lk}(k';z) e^{-in\phi_1}\delta(k'-|\boldsymbol{k}_1^\perp|)$$

and

$$C(\boldsymbol{k}_2) = \sum_{k'',m}\int dz e^{-ik_2^z z}\,c_m(k'';z)\,e^{im\phi_2}\delta(k''-|\boldsymbol{k}_2^\perp|)$$

where $\phi_1 = \phi_{\boldsymbol{k}_1^\perp}$ and $\phi_2 = \phi_{\boldsymbol{k}_2^\perp}$. By direct substitution in the expression for $Q_a^x(l,k)$ we can find

$$Q_a^x(l,k) = \sigma \sum_{k',n}\sum_{k'',m} c_m(k'';a) d_n^{*lk}(k';a) \int_0^{2\pi} d\phi_1 \int_0^{2\pi} d\phi_2 \int_0^\infty |\boldsymbol{k}_1^\perp| d|\boldsymbol{k}_1^\perp| \int_0^\infty |\boldsymbol{k}_2^\perp| d|\boldsymbol{k}_2^\perp|$$
$$\times \frac{\delta(k'-|\boldsymbol{k}_1^\perp|)\delta(k''-|\boldsymbol{k}_2^\perp|)(k_1^x-k_2^x)e^{im\phi_2}e^{-in\phi_1}}{|\boldsymbol{k}_1^\perp - \boldsymbol{k}_2^\perp|^2 + q_{\Delta E}^2}$$

which becomes, after exploiting the properties of Dirac delta functions:

$$Q_a^x(l,k) = \sigma \sum_{k',n}\sum_{k'',m} c_m(k'';a) d_n^{*lk}(k';a) k' k'' \int_0^{2\pi} d\phi_1 \int_0^{2\pi} d\phi_2 \frac{(\tilde{k}_1^x - \tilde{k}_2^x)e^{im\phi_2}e^{-in\phi_1}}{\left|\tilde{\boldsymbol{k}}_1^\perp - \tilde{\boldsymbol{k}}_2^\perp\right|^2 + q_{\Delta E}^2}$$

where $\tilde{\boldsymbol{k}}_1^\perp = k'(\cos\phi_1, \sin\phi_1)$ and $\tilde{\boldsymbol{k}}_2^\perp = k''(\cos\phi_2, \sin\phi_2)$.

To proceed further with an analytical treatment we need to introduce two approximations:

1. the integrand $\frac{1}{\left|\tilde{\boldsymbol{k}}_1^\perp - \tilde{\boldsymbol{k}}_2^\perp\right|^2 + q_{\Delta E}^2}$ has its larger values once $\left|\tilde{\boldsymbol{k}}_1^\perp\right| \approx \left|\tilde{\boldsymbol{k}}_2^\perp\right|$, i.e. if $k' \approx k''$. Therefore in the following we will fix $k'$ equal to $k''$;
2. we consider only the $m = 0$ component of the incident beam ($c_{m=0}(k'';a)$) in the sum appearing in the expression for $Q_a^x(l,k)$ as other components only come from neighbouring columns.

Under these assumptions we can write

$$Q_a^x(l,k) \approx \sigma \sum_{k',n} c_0(k';a)\,d_n^{*lk}(k';a) k'^3 \int_0^{2\pi} d\phi_1 \int_0^{2\pi} d\phi_2 \frac{(\cos\phi_1 - \cos\phi_2)e^{-in\phi_1}}{4k'^2 \sin^2\left(\frac{\phi_2-\phi_1}{2}\right) + q_{\Delta E}^2}$$

Now we group the integrals over $\phi_1$ and $\phi_2$ in a single function $\zeta(k')$ defined as

$$\zeta(k') = \int_0^{2\pi} d\phi_1 \int_0^{2\pi} d\phi_2 \frac{(\cos\phi_1 - \cos\phi_2)e^{-in\phi_1}}{4k'^2 \sin^2\left(\frac{\phi_2-\phi_1}{2}\right) + q_{\Delta E}^2}$$

Performing the substitution $\phi_2 - \phi_1 = \tau$ we find

$$\zeta(k') = \int_0^{2\pi} d\phi_1 \cos\phi_1\, e^{-in\phi_1} \int_{-\phi_1}^{2\pi-\phi_1} d\tau \frac{1}{4k'^2 \sin^2\left(\frac{\tau}{2}\right) + q_{\Delta E}^2}$$
$$- \int_0^{2\pi} d\phi_1 \cos\phi_1\, e^{-in\phi_1} \int_{-\phi_1}^{2\pi-\phi_1} d\tau \frac{\cos\tau}{4k'^2 \sin^2\left(\frac{\tau}{2}\right) + q_{\Delta E}^2}$$
$$+ \int_0^{2\pi} d\phi_1 \sin\phi_1\, e^{-in\phi_1} \int_{-\phi_1}^{2\pi-\phi_1} d\tau \frac{\sin\tau}{4k'^2 \sin^2\left(\frac{\tau}{2}\right) + q_{\Delta E}^2}$$

All the integrands of the integrals over $\tau$ are functions $f$ such that $f(\tau) = f(\tau - 2\pi)$: therefore

$$\int_{-\phi_1}^{2\pi-\phi_1} d\tau f(\tau) = \int_{-\phi_1}^{0} d\tau f(\tau) + \int_{0}^{2\pi-\phi_1} d\tau f(\tau) = \int_{-\phi_1}^{0} d\tau f(\tau) + \int_{0}^{2\pi} d\tau f(\tau) - \int_{2\pi-\phi_1}^{2\pi} d\tau f(\tau)$$

By the substitution $\tau \to \tau - 2\pi$ in the last integral we can easily find

$$\int_{2\pi-\phi_1}^{2\pi} d\tau f(\tau) = \int_{-\phi_1}^{0} d\tau f(\tau)$$

which simplifies with the first term: so the dependence of the overall integral from $\phi_1$ disappears. Therefore

$$\zeta(k') = \int_0^{2\pi} d\phi_1 \cos\phi_1\, e^{-in\phi_1} \left[ \int_0^{2\pi} d\tau \frac{1}{4k'^2 \sin^2\left(\frac{\tau}{2}\right) + q_{\Delta E}^2} - \int_0^{2\pi} d\tau \frac{\cos\tau}{4k'^2 \sin^2\left(\frac{\tau}{2}\right) + q_{\Delta E}^2} \right]$$

Solving the integral over $\phi_1$ and defining

$$A(k') = \frac{1}{2}\left[ \int_0^{2\pi} d\tau \frac{1}{4k'^2 \sin^2\left(\frac{\tau}{2}\right) + q_{\Delta E}^2} - \int_0^{2\pi} d\tau \frac{\cos\tau}{4k'^2 \sin^2\left(\frac{\tau}{2}\right) + q_{\Delta E}^2} \right]$$

we obtain

$$Q_a^x(l,k) \propto \sigma \sum_{k',n} c_0(k';a)\, d_n^{*lk}(k';a) k'^3 A(k') [\delta_{n,1} + \delta_{n,-1}] \quad (B1)$$

From which it is easy to understand that only the $n = \pm 1$ components of the back-propagated Bessel beam effectively contribute to $Q_a^x(l,k)$.

# References

[1]  P. Schattschneider, S. Rubino, C. Hebert, J. Rusz, J. Junes, P. Kovak, E. Carlino, M. Fabrizioli, G. Panaccione and G. Rossi, Nature **441**, 486 (2006).
[2]  D. Song, Z. Wang and J. Zhu Ultramicroscopy, 201, 1-17, (2019).
[3]  Z. Wang, A. H. Tavabi, L. Jin, J. Rusz, D. Tyutyunnikov, H. Jiang, Y. Moritomo, J. Mayer, R. E. Dunin-Borkowski, R. Yu, J. Zhu, X. Zhong, Nature Materials **17**, 221–225 (2018)
[4]  P. Schattschneider, B. Schaffer, I. Ennen, and J. Verbeeck, Phys. Rev.B **85**,134422 (2012)
[5]  J. Rusz, S. Rubino and P. Schattschneider, Phys. Rev. B **75**, 214425 (2007).
[6]  P. Schattschneider, C. Herbert, S, Rubino, M. Stoger-Pollach, J. Rusz, P. Novak, Ultramicroscopy **108** 433-438 (2008).
[7]  J. Verbeeck, H. Tian and P. Schattschneider, Nature **467**, 301-304 (2010).
[8]  M. Uchida and A. Tonomura, Nature **464**, 737-739 (2010).
[9]  B.J. McMorran, A. Agrawal, I.M. Anderson, A.A. Herzing, H.J. Lezec, J.J. McClelland and J. Unguris, Science **331**, 192-195 (2011).
[10] D. Pohl, S. Schneider, P. Zeiger, J. Rusz, P. Tiemeijer, S. Lazar, K. Nielsch and B. Rellinghaus, Scientific Reports **7**, 934 (2017).
[11] A. Béché, R. Juchtmans and J. Verbeeck, Ultramicroscopy **178**, 12-19, (2017)
[12] R. Van Boxem, B. Partoens and J. Verbeeck, Phys. Rev. A **91**, 032703 (2015).
[13] J. Rusz and S. Bhowmick, Phys. Rev. Lett. **111**, 105504 (2013).
[14] J. Rusz, J-C. Idrobo and S. Bhowmick, Phys. Rev. Lett. **113**, 145501 (2014).
[15] J. Rusz, J-C. Idrobo and L. Wrang, Phys. Rev. B **94**, 144430 (2016).
[16] J. Rusz, S. Bhowmick, M. Eriksson, and N. Karlsson Phys. Rev. B , **89**, 134428 (2014)
[17] J.C. Idrobo, J.Rusz, J. Spiegelberg, M.A. McGuire, C.T. Symons, R.R.Vatsavai, C.Cantoni and A.R. Lupini Adv Struct Chem Imag 2:5 (2016)
[18] E. J. Kirkland, Advanced Computing in Electron Microscopy, 2$^{nd}$ ed. (Springer, New York, 2010).
[19] S. Loffler and P. Schattschneider, Acta Cryst. A 68, 443–447 (2012).
[20] B.G. Mendis, Ultramicroscopy 149, 74-85(2015).
[21] T. R. Harvey, V. Grillo, and B. J. McMorran, Microsc. Microanal. 22 (Suppl 3), 1708 (2016)
[22] T. Schachinger, S. Löffler, A. Steiger-Thirsfeld, M. Stöger-Pollach, S. Schneider, D. Pohl, B. Rellinghaus and P. Schattschneider, Ultramicroscopy, **179**, 15-23 (2017).
[23] V. Grillo, A.H. Tavabi, F. Venturi, H. Larocque, R. Balboni, G.C. Gazzadi, S. Frabboni, P.H. Lu, E. Mafakheri, F. Bouchard, R.E. Dunin-Borkowski, R.W. Boyd, M.P.J. Lavery, M.J. Padgett and E. Karimi, Nature Communications **8**, 15536 (2017).
[24] B.J. McMorran, T.R. Harvey and M.P.J. Lavery, New J. Phys. **19**, 023053 (2017).
[25] G.C.G. Berkhout, M.P.J. Lavery, J. Courtial, M. W. Beijersbergen and M.J. Padgett, Phys. Rev. Lett. **105**, 153601 (2010).
[26] M. Zanfrognini, E. Rotunno, S. Frabboni, A. Sit, E. Karimi, U. Hohenester and V. Grillo, ACS Photonics 6, 620 (2019).
[27] I. Ennen, S.Loffler, C.Kubel, D.Wang, A.Auge, A.Hutten, and P.Schattschneider J. Magn. Mag. Mater., 324 2723 (2012)
[28] J.Rusz, S.Muto, J. Spiegelberg, R. Adam, K.Tatsumi, D. E. Burgler, P.M.Oppeneer and C.M. Schneider, Nat. Commun., **7**, 12672 (2016)
[29] H. Ali, T. Warnatz, L. Xie, B. Hjörvarsson, and K. Leifer, Ultramicroscopy, 196, 192-196, (2019)
[30] H. Ali, D. Negi, T. Warnatz, B. Hjörvarsson, J. Rusz, K. Leifer, https://arxiv.org/abs/1908.11755, (2019)

[31] T. Thersleff, L. Schönström, C-W Tai, R. Adam, D. E. Bürgler, C. M. Schneider, S. Muto, J. Rusz, https://arxiv.org/abs/1908.09132 (2019)

[32] J. Rusz, A. Lubk, J. Spiegelberg and D. Tyutyunnikov, Phys. Rev. B **96**, 245121 (2017).

[33] C. Dwyer, Ultramicroscopy **104**, 141 (2005).

[34] J. Verbeeck, P. Schattschneider and A. Rosenauer, Ultramicroscopy **109**, 350 (2009).

[35] H. Kohl and H. Rose, Adv. Electron. Electron Phys. **65**, 173 (1985).

[36] J. Rusz, S. Rubino, O. Eriksson, P.M. Oppeneer and K. Leifer, Phys.Rev.B **84**, 064444 (2011).

[37] S. Löffler, V.Motsch, and P.Schattschneider Ultramicroscopy, 131, 39-45 (2013)

[38] J. Rusz, Ultramicroscopy **177**, 20-25 (2017).

[39] J. Rusz, O. Eriksson, P. Novák, P. M. Oppeneer, Phys. Rev. B 76, 060408, (2007)

[40] L. Calmels, F. Houdellier, B. Warot-Fonrose, C. Gatel, M. J. Hÿtch, V. Serin, E. Snoeck, and P. Schattschneider J.M. Auerhammer and P. Rez Phys. Rev. B, **40**, 4 (1989)

Phys. Rev. B 76, 060409 (2007)

[41] J.M. Auerhammer and P. Rez Phys. Rev. B, **40**, 4 (1989)

[42] Kirkland E. J., Advanced Computing in Electron Microscopy (Springer, New York, 2010) 2nd ed

[43] V. Grillo, E. Rotunno Ultramicroscopy 125, 97-111 (2013)

[44] E. Rotunno, A.H. Tavabi, E. Yucelen, S. Frabboni, R.E. Dunin Borkowski, E. Karimi, B.J. McMorran, and V. Grillo, Phys. Rev. Applied 11, 04407 (2019).

[45] D.S. Negi, J.C. Idrobo and J. Rusz, Scientific Reports **8**, 4019 (2018).

[46] P. Blaha, K. Schwarz, G.K.H. Madsen, D. Kvasnicka, J. Luitz, R. Laskowski, F. Tran and L.D. Marks, WIEN2k, An Augmented Plane Wave + Local Orbitals Program for Calculating Crystal Properties (Karlheinz Schwarz, Techn. Universität Wien, Austria), 2018. ISBN 3-9501031-1-2.

[47] J.P. Perdew, K. Burke and M. Ernzerhof, Phys. Rev. Lett. **77**, 3865 (1996).

[48] M. Mirhosseini, M. Malik, Z. Shi and R.W. Boyd, Nature Communications **4**, 2783 (2013).